\documentclass[conference]{IEEEtran}
\IEEEoverridecommandlockouts
\usepackage{amsmath,amssymb,amsfonts}
\usepackage{color}
\usepackage{algorithmic}
\usepackage{slashbox}
\usepackage{array}
\usepackage{pifont}
\usepackage{graphicx}
\usepackage{textcomp}
\usepackage[utf8]{inputenc}
\usepackage{graphicx}
\usepackage{hyperref}
\usepackage{amsmath}
\usepackage{amsfonts}
\usepackage{amssymb}
\begin{document}
\title{Analysis and Improvement of a Lightweight Anonymous  Authentication Protocol for Mobile Pay-TV Systems (Full text) \protect\thanks{The summary of this article has been submitted at \url{http://ist2018.itrc.ac.ir/} $^*$Corresponding author: Saeed Banaeianfar saeed\_ banaeian\_ far@yahoo.com}} 
\author{\IEEEauthorblockN{1\textsuperscript{st} Saeed Banaeian Far$^*$}
\IEEEauthorblockA{\textit{Department of Electrical and Computer Engineering,}\\ \textit{Science and Research Branch} \\
\textit{Islamic Azad University}\\
Tehran, Iran \\
Saeed.banaeian@srbiau.ac.ir}
 \and 
\IEEEauthorblockN{2\textsuperscript{nd} Mahdi R.Alagheband}
\IEEEauthorblockA{\textit{Department of Electrical and Computer Engineering,}\\ \textit{Science and Research Branch} \\
\textit{Islamic Azad University}\\
Tehran, Iran \\
m.alaghband@srbiau.ac.ir}}

\maketitle

\begin{abstract}
For many years, the pay-TV system has attracted a lot of users. Users have recently expressed the desire to use mobile TV or mobile payment via anonymous protocols. The mobile users have also received their services over cellular communications networks. Each mobile device receives services from each head end systems. With increasing numbers of users and the expansion of Internet, user's privacy has become crucial important. When a device leaves the head end system's range, it must receive services from another head end system. In this paper, we review Chen \textit{et al}'s scheme and we highlight some weaknesses, including \textit{privilege insider attack} and \textit{user traceability attack}. Finally, we alleviate the scheme and analyze the alleviated scheme using both heuristic and formal methods.
\end{abstract}
\section{Introduction}
After  World War $II$, wireless communications were launched, and mobile services gradually became available. At that time, there was only one mobile operator that provided service to a few users. Then, the second generation of mobile communications was introduced as cellular networks under the $GSM$ standard \cite{rani}. Mobile communications rapidly progressed. To date, communications have changed significantly four times. These changes and technological mutations were introduced as different generations of wireless communications technology so that today, the fourth generation of this method of communication is utilized. It is predicted that a new generation of mobile communications will be introduced in 2020 that will provide users with great speed and accuracy\cite{rani,bhmu,bki}.
Pay-TV system has attracted many users for almost $30$ years. In 1994, the number of people who used this technology reached $3.45$ million in England. It doubled after 4 years \cite{marar}. Currently, numerous users use mobile devices to watch TV, and many communication systems have been provided for using the mobile-TV services \cite{marar,liu,mosa,aubr}. In these systems, the user can receive services after registration in a head end system ($HES$) network and store his/her information in the database server ($DBS$) of the $HES.$ At first, the $HES$ only broadcasts one authentication message to all the users who request the same service \cite{liu}. Additionally, in \cite{aubr}, a user can access a television channel and play any video on his/her mobile phone.\\
Since smart card-based networks and mobile phone users are  shifting to an ad hoc and comletely mobile mode, $HES$ cannot provide service to users everywhere. Therefore, when mobile users leave an area covered by a $HES$, they should receive services from another $HES$ \cite{shabak,AIHC1,copo,loya}. First of all, they have to be authenticated again. In \cite{shabak}  the MobiCash protocol based on elliptic curve cryptography ($ECC$)  was proposed, and \cite{AIHC1} used symmetric encryption functions. Constantin Popescu and  Lo-yao Yeh's  schemes are also based on bilinear pairing \cite{copo,loya}.\\
Recently, user privacy has acquired  special significance so that the demand for anonymous communication in networks has increased, and service providers have to authenticate users remotely and anonymously \cite{we,guya,zhan}. Bapana describes  an anonymous authentication protocol suitable for distributed computer networks \cite{surba}. Yang \textit{et al.} proposed a two-party secure roaming protocol based on identity based signatures ($IBS$) \cite{guya}. There are multiple servers, and each server manages a set of subscribers who are users of the network. De-dong \textit{et al.} presented a model of two access modes: self-access and cross-domain access \cite{zhan}. In self-access, the internet service providers ($ISP$) provide service to users directly, and cross-domain access is similar to that of a roaming network. \\
Anonymous authentication schemes could meet these requirements. The validity and legality of a user's identity is approved in anonymous authentication schemes while divulging their true identity to no one. In some schemes, not only is the user's identity anonymous on public networks and channels, but  users inside the network and attackers also cannot retrieve the user's $ID$ \cite{surba,chen}. Even the server occasionally does not realize the user's $ID$ \cite{chit}. In some schemes, there is no registration table for user authentication on the server \cite{chit}, but the server can validate and authenticate the users anonymously. \\
Our main contributions are listed follows:
\begin{itemize}
\item We present a great user anonymity. It means that we assume that an adversary has the server secret key and user password, but it cannot obtain the user's identity.
\item We improve the Chen's scheme \cite{chen} against privilege insider attack attack. It means that the severs have no ability to obtain the users sensitive information.
\item We improve the Chen's scheme \cite{chen} against user traceability attack.
\item We analyze the alleviated scheme using both heuristic and formal methods.
\end{itemize}

In this paper, after discussing related work in section $2$ and prerequisites for the scheme in section $3$, we describe and analyze Chen \textit{et. al}'s scheme\cite{chen} in section $4.$ It should be noted that Kim \textit{et al} have also promoted this scheme against stolen  smart card-based attacks \cite{kim}, but it is almost infeasible as long as the user chooses just strong password because it requires succeed brute force attack. In the following, we explain our alleviated scheme. Then, we analyze the alleviated scheme in  heuristic and formal forms. Finally, we compare our scheme with recently proposed schemes.

\section{Related work}

In this section, we categorize pay-TV systems in $4$ groups. At first, we describe signature-based pay-TV systems. Blind signature is practical way to verify user authentication by valid party, anonymously. Other categories are based on bilinear pairing, digital signature, encryption/decryption function, and hash function. Since hash function are light and secure, it is applied for lightweight schemes which are suitable for weak devices such as sensors and smart phones. We depict this section on the table \ref{rw}, in summery.


\begin{table*}
\caption{An overview on the section $2$}
\centering
\resizebox{\textwidth}{!}{
\begin{tabular}{|c|c|c|c|c|c|l|}
\hline
 \textbf{Information $\rightarrow$} &  & \textbf{Base Article or} & \multicolumn{3}{c|}{\textbf{Cryptographic functions}} & \textbf{Contribution} \\
 \textbf{Schemes} $\downarrow$ & \textbf{Year} & \textbf{improvement of.} & $BP$ & $E/D/Sig$ & $H$ & (in summery)  \\
\hline \hline
Camenicsh \cite{jancam} & 1996 & Brickell 1995 \cite{c3} & No & Yes & Yes & Proposed an anonymous electronic\\
  &  & Stadler 1995 \cite{c16} &  &  &  & payment system  \\ \hline
Bakhtiari \cite{shabak} & 2009 & Abbadasari 2004 \cite{b9} & No & Yes & Yes & Proposed a MobiCash system \\ \hline
Chen \cite{chen} & 2011 & Yang 2009 \cite{yc} & No & No & Yes & Improved Yang's scheme \cite{yc} against \\ 
  &  &  &  &  &  & insider and impersonation attack \\ \hline 
Kim \cite{kim} & 2012 & Chen 2011 \cite{chen} & No & No & Yes & Improved Chen's scheme \cite{chen} against smart \\
  &  &  &  &  &  & card-based attacks. We explain in the \\
  &  &  &  &  &  & section 4 that the improvement is not true \\ \hline
Wang \cite{AIHC4} & 2012 & Sun 2009 \cite{wangimp} & Yes & Yes & Yes & Improved Sun's scheme \cite{wangimp} against\\
  &  &  &  &  &  & MitM and impersonation attacks \\ \hline
Liu \cite{liu} & 2013 & Sun 2009 \cite{wangimp} & Yes & Yes & Yes & Privacy preserving \\
  &  &  &  &  &  & Reduce the computation overhead \\ \hline
Tsai \cite{jilu} & 2014 & Li 2012 \cite{xili} & No & No & Yes & Proposed a protocol based on chaotic map \\ \hline
Sabzinejad  & 2014 & Yeh 2012 \cite{loya} & Yes & Yes & Yes & Improved Yeh's scheme \cite{loya} against head-\\
 Farash \cite{mosa} &  &  &  &  &  & end system impersonation attack \\
  &  &  &  &  &  & Reduced the computation overhead \\ \hline
Heydari \cite{h2015} & 2015 & Wang 2012 \cite{AIHC4} & Yes & Yes & Yes & Improved Wang's scheme \cite{AIHC4} against \\
  &  &  &  &  &  & impersonation attack \\
  &  &  &  &  &  & Reduced the computation overhead \\ \hline
Kou \cite{wech} & 2015 & Choi 2014 \cite{k4} & No & No & Yes & Improved Choi's scheme \cite{k4} against \\
  &  &  &  &  &  & stolen smart card and impersonation attacks \\ \hline
Wu \cite{AIHC1} & 2016 & He 2016 \cite{AIHC3} & Yes & Yes & Yes & Added anonymity to He's scheme \cite{AIHC3} \\ \hline
Wu \cite{new4} & 2017 &  Sabzinejad  &  &  &  & Added mutual authentication to\\
  &  & Farash 2014 \cite{mosa} &  &  &  & Sabzinejad Farash \cite{mosa} \\ \hline
Arshad \cite{ar2017} & 2017 & Wang 2012 \cite{AIHC4} & No & Yes & Yes & Made Wang's scheme \cite{AIHC4} efficient, and \\
 &  &  &  &  &  & implements it on $FPGA$ boards \\ \hline
Biesmans \cite{b2018} & 2018 & Attrapadung 2009 \cite{b14} & Yes & Yes & Yes & Privacy preserving \\ \hline \hline
\multicolumn{7}{|l|}{\textbf{Note:}}\\
\multicolumn{2}{|l}{$BP$: Bilinear Pairing} & \multicolumn{4}{l}{$E/D/Sig$: Encryption/Decryption/Digital Signature} & \multicolumn{1}{l|}{$H$: Hash function} \\
\hline
\end{tabular}
}
\label{rw}
\end{table*}

\begin{itemize}

\item \textit{Pairing-based pay-TV systems}\\ Other researchers proposed heavy scheme based on bilinear pairnig, such as Wang \textit{et al} proposed an authentication scheme for access control in mobile pay-TV systems, in 2012 \cite{AIHC4}. Their protocol was resist against forgery, Man-in-the-Middle $MitM$, and replay attacks. Thus, an adversary can pass the verification phase successfully. Its performance is good, but as we said, it is based on bilinear pairing and that's not suitable for lightweight devices. In 2013, Liu used identity-based encryption in his scheme \cite{liu}. A number of schemes also used  cryptographic functions and bilinear pairings (e.g., \cite{mosa,AIHC1,copo,loya}). Sabzinejad Farash made improvments \cite{loya} against impersonation attacks (User impersonation and $HES$ impersonation) in \cite{mosa}. Sabzinejad Farash's scheme \cite{mosa} is also a robust and secure system which is the running time of protocol shorter than the previous schemes. However, his proposed scheme is designed with bilinear pairing. It is too heavy and unsuitable for the weak devices. In 2015, Heydari \textit{et al} proposed an authentication scheme resists against impersonation attack. They launched their attack on issue phase and generalized it on other phases \cite{h2015}. \\ Recently, Wu et al. proposed an authentication schemes for mobile pay-TV, but it does not support anonymity \cite{AIHC1}. Anonymity is being supported through pairing transform in \cite{AIHC3}. Also, in 2017, Wu \textit{et al.} proved that the Sabzinejad protocol has some weaknesses \cite{new4}. For example, it does not support mutual authentication. But there is no modified scheme. Then, Biesmans \textit{et al} proposed pay-per-view and a pay-per-channel that protect users' privacy \cite{b2018}.

\item \textit{Signature-based pay-TV systems}\\ A user's connection with banks is another payment method that can be mentioned. Blind signature is another method for anonymization. In this method, a legitimate party signs the blinded message of users. After signing, other people can see the original message along with the valid signatures of the legal party. In 1996, Camenisch presented a communication scheme in an anonymous way in which a blind signature was used \cite{jancam}. Subsequently, authors have tried to provide more efficient schemes for mobile-pay systems. Customers in this scheme have to open an anonymous account and there is no need for the bank to identify the customers. In 2009 Bakhtiari \textit{et al} presented the $MobiCash$ scheme based on the blind signature and the customer's relationship with the bank \cite{shabak}. Its blind signature is based on $ECDSA$ crypto system. In $2016$, Wu \textit{et.al} proposed an efficient scheme \cite{AIHC1}. Their scheme is a powerful scheme based on user signature.  The user who wants to uses TV, must register as legal user via proposed scheme and creates session key to watch the TV. The user signs its message and sends it to server. Server verifies the received message with $open.algorithm$ which is proposed in this scheme.

\item \textit{Encryption-based pay-TV systems}\\ Encryption-based schemes are most practical classification. This category neither heavy nor light. Thus, are not suitable for weak devices. For example, Yang J-H and Chang proposed $ID$-based scheme on $ECC$ crypto system for remote user authentication \cite{yc}. It has some drawbacks such as vulnerability to insider and impersonation attacks \cite{chen}. In the hash-based pay-TV system (the next category), we describe Chen's scheme \cite{chen} and then analyse it in the section $4$. In $2017$ Arshad \textit{et al} proposed an efficient scheme \cite{ar2017} based on Wang's scheme \cite{AIHC4}. But, there is no bilinear pairing functions and Arshad's scheme is  easy to implement on $FPGA$ boards.

\item \textit{Hash-based pay-TV systems}\\ In 2011, Chen modified the $ECC$-based scheme of Yang J-H and Chang \cite{yc}. The modified protocol is lightened and redesigned only with hash function but without $ECC$ using \cite{chen}. Then in 2012, Kim \textit{et al}  improved Chen's scheme \cite{chen} and made it robust against smart card-based attacks \cite{kim}. We  will explain in section $4$ that the improvement is not true.\\ In recent years, low energy consumption on smart cards has been motivated by an increase in the energy efficiency and productivity of schemes so that some of the designed schemes for smart cards are hash-based and have high energy consumption efficiency (e.g., \cite{xili,jilu,wech}). In 2014, Tsai proposed a light anonymous authentication protocol \cite{jilu}. Then, in 2015, Kuo \textit{et al.}  also presented a lightweight scheme based on smart cards \cite{wech}. The lightweight schemes are popular, since light devices have been developed. So, we focus on this category of pay-TV systems.
\end{itemize}

In the section $5$, we discuss the \cite{mosa,loya,chen,kim} and our scheme, in the compare them with our improved scheme. We illustrate that our alleviated scheme is more secure than noted schemes.

\section{Preliminaries}
In this section, we explain preliminaries of our paper. These functions are used for lightweight protocols with low power consumption, so common encryption functions are not used. After presenting the required security features in anonymous authentication schemes, we briefly explain cellular communication. Finally, we explain the analysis of schemes through a formal method. 

\subsection{Parameters and Entities Description}

\paragraph{\textbf{Describe the entities involved in this paper}}
In Table \ref{notations}, the list of entities and parameters are depicted. Below, we explain the role of each of the entities \cite{chen}.


\begin{table*}
\caption{List of notations}
\centering
\begin{tabular}{|cl|cl|}
\hline
\textbf{Entities} & \textbf{Description} & \textbf{Parameters} & \textbf{Description} \\
\hline \hline
 $HES$ & Head End System          & $S$ & Server \\
 $SAS$ & Subscriber Authorization System           & $U_{i}$ & The i\textit{th} user \\
 $SMS$ & Subscriber Management System          & $ID_{i}$ & $ID$ of i\textit{th} user \\
 $CW$  & Control Word & $PW_{i}$        & Password of i\textit{th} user \\
 $CAS$ & Conditional Access System         & $b$ & Random number \\
 $ECM$ & Entitlement Control Message        & $N$ & User registration number \\
 $EMM$ & Entitlement Management Message        & $T$ & Time stamp \\
 $DBS$ & Data Base Server          & $\Delta T$ & $T_{i}-T_{j}$ \\
 $MUX$ & Multiplexer            & $\Theta$ & Token for issue phase  \\
 $DEMUX$ & Demultiplexer            & $\gamma$ & Token for subscription phase \\
 $TX$ & Transmitter            & $\gamma_{i}$ & Token for hand-off phase \\
 $RX$ & Receiving module          & $h(.)$ & Secure one-way hash function \\
 $MS$ & Mobile device            & $\oplus$ & XOR operation \\
 $DVB$ & Digital Video Broadcast          & $\Vert$ & Concatenate operation \\
 $\mathcal{A}$ & The Adversary          & $y$ & The secret key of the remote server\\
  &  & $*$ & The stared parameters are generated by adversary \\
\hline
\end{tabular}
\label{notations}
\end{table*}


\begin{itemize}
\item \textbf{HES:} A system sending broadcast TV service to receivers.
\item \textbf{Receiver: }A mobile device with a $CAS$ module used for access control.
\item \textbf{SAS/SMS:} Subsystems responsible for subscriber authorization and management.
\item \textbf{Encrypter/Decrypter:} Components for encrypting and decrypting CW, keys and sensitive information.
\item \textbf{Multiplexer/Demultipexer:} Components for multiplexing and demultiplexing A/V, data or IP to MPEG-2.
\item \textbf{Scrambler/Desclamber:} Components for signal scrambling and the reverse engineering of Scrambler.
\item \textbf{TX/RX:} Subsystems for signal transmission and receiving.
\item \textbf{ECM/EMM:} Defined by $DVB$ as two conditional access messages.
\end{itemize}


\subsection{Security Requirements}

In this section, we mention the definitions of security requirements and the need for anonymous authentication protocols with  multi-server service providers. Noted that the hash function have to be secure in standard model against relevant attacks. One-way hash functions with no collisions are functions with variant input and constant length output. From their characteristics, we can note that they do not have collisions and that they are one-way \cite{nanal}.

\begin{itemize}

\item \textbf{Privacy Preserving}\\
Privacy is a range of personal and private information of the user that the user wants to be protected and unavailable to $\mathcal{A}$s \cite{yand}. In this paper, user identity requires protection. Because users want to $log-in$ anonymously and keeps his/her identity private.

\item \textbf{User Anonymity}\\
User anonymity is a kind of privacy policy in networks. User anonymity means that user's identities cannot be obtained and find a link to trace the users by any channel eavesdropping, stolen smart card, or access to the user database stored in server memory \cite{link}. According to increasing user requests to join the networks and uses internet-based services, user's privacy has become particularly important and identity anonymity is more considered.

\item \textbf{User Traceability}\\
Traceability means that if a user $logs-in$ to a server several times, or to multiple servers in several different points, $\mathcal{A}$ or other users cannot determine wheter is the same user that was previously $logged-in$ to the server or not \cite{andr}.

\item \textbf{Resistance against privilege insider attack}\\
There are many $HES$es in the cellular network and users can get services from them. They authenticate users and then the users can use the services. To authenticate users, the $HES$es obtain the real users' identity and then verify their $log-in$ request. It is clearly that in this attack all $HES$es know the real users' identity and if one of the $HES$es  is malicious, the users' privacy is broken \cite{2018sd}. But, we want to the $HES$es learn no privacy information about the users' identity.

\item \textbf{Forward and Backward Security}\\
Forward security means that when the user is out of the network (or network service is revoked) and he is not a member of the network he must not retrieve encrypted messages after leaving or revocation. In fact, it means that the set of keys in the next sessions must be independent of the set of keys in the previous ones. Backward security means that if a user recently was a member of that network with a new key to server, this user would no longer retrieve previous session keys. This user cannot retrieve the previous encrypted information  by having either the exchanged information in the past or the current key \cite{bend}. 

\item \textbf{Mutual Authentication}\\
For secure communication, it is necessary that both parties presuade each other to confrim the identity. So, the user is known to the server, and the user is able to authenticate the server through the mutual authentication protocol \cite{chit}.

\end{itemize}

\subsection{Formal Security Analysis}
Many proposed anonymous authentication protocols have been analyzed via ad hoc methods, but all of their drawbacks have not been discovered. Hence, there is no doubt that a formal method to discover the privacy and security drawbacks is required.  $\mathcal{A}$'s capabilities and threat models are classified in formal analysis. In this case, the adversary is capable of not only eavesdropping on the channel but also revealing secret data via data recovery through smart card power analysis attack \cite{dabt}.\\
A game-based model is applied to prove each attack. $\mathcal{A}$ tries to success in the designed game. We illustrate that $\mathcal{A}$  succeeds in designed game over Chen's scheme \cite{chen}. However, it fails in designed games over our improved scheme.
According to the protocol's attributes, a formal analysis method has three functions \cite{dabt,yand}: (i) the experiment function, (ii) the success function, and (iii) the probability function. as follows:

\begin{itemize}
\item \textbf{Experiment function (EXP):} $\mathcal{A}$ performs the process to get the required information.
\item \textbf{Success function (Succ):} It specifies how successful $\mathcal{A}$ is in gaining important data.
\item \textbf{Probability function (Pr):}  $\mathcal{A}$'s probability of success in the recovery of secret values.
\end{itemize}

If the probability of success is negligible ($\epsilon$), the latter protocol is secure against assumed $\mathcal{A}$ \cite{yand}. $$Succ_{\mathcal{A}}^{Protocol-name}=Pr[EXP_{\mathcal{A}}^{H.P.}]\leq \epsilon$$

\subsection{Adversary abilities}
In this section we describe $\mathcal{A}$ abilities. We allow $\mathcal{A}$ to achieve all parameters stored in smart card and database of servers, and it can eavesdrop the public channel, to show the security power of our alleviated scheme.\\
In the following we describe $\mathcal{A}$ abilities \cite{we,2018sd}, briefly:
\begin{itemize}
\item $\mathcal{A}$ can eavesdrop the public channel. 
\item $\mathcal{A}$ can achieve to parameters stored in smart card.
\item $\mathcal{A}$ can achieve to verification table which is the servers has access to it.
\end{itemize}
In the section $4.3$, we show our scheme is secure against all of smart card-based and stolen server attacks, as well as privacy drawbacks such as lack of anonymity and traceability.
In the following, after explaintion of Chen's scheme \cite{chen} and its weaknesses, we depirt our modified scheme indetails.

\section{Review of the Chen et al. Scheme}
\begin{figure*}
\centering
\includegraphics[scale=0.15]{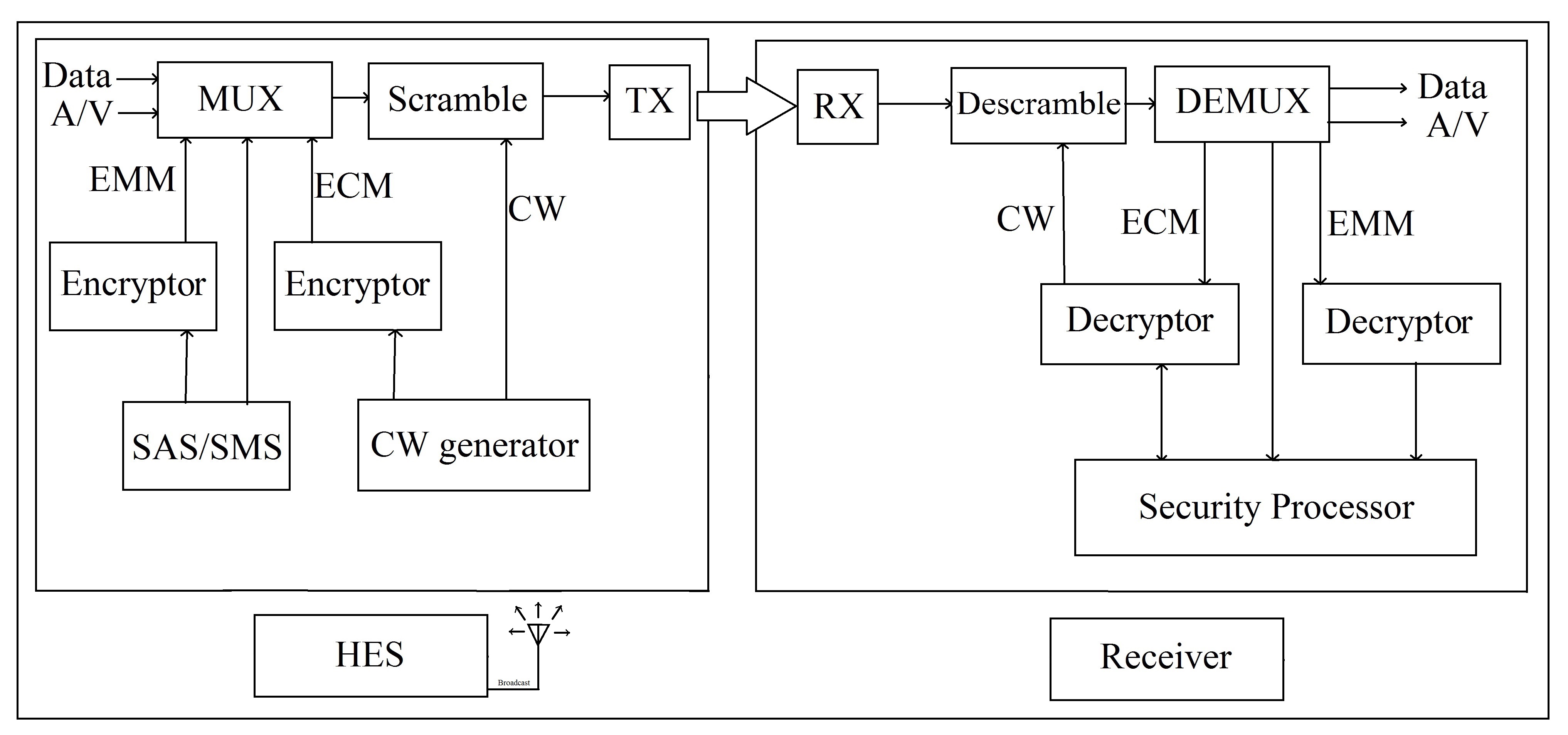}
\caption{The structure of $CAS$ in a general mobile pay-TV system \cite{chen}}
\label{shekl e chen}
\end{figure*}

\begin{figure*}
\centering
\includegraphics[scale=0.15]{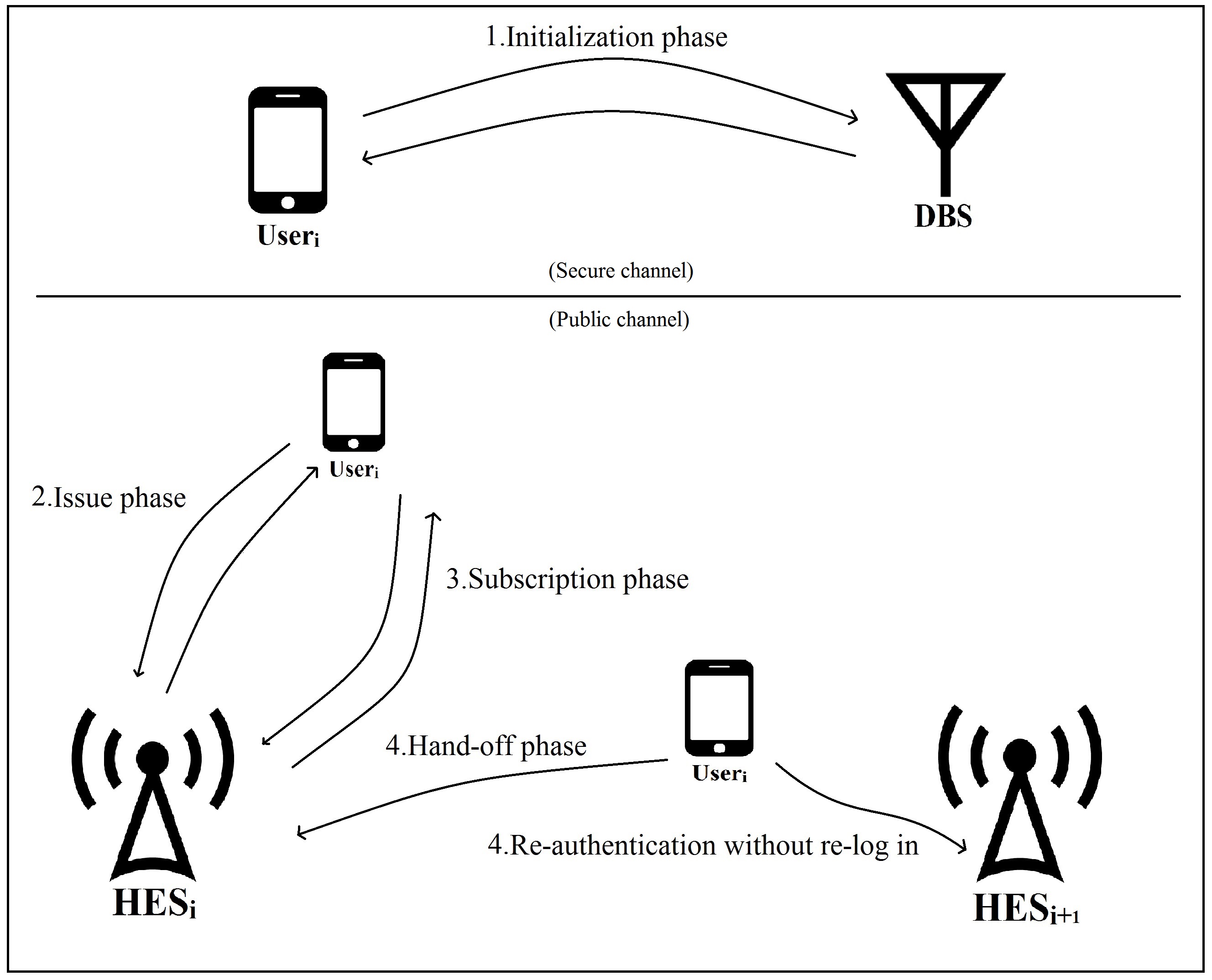}
\caption{The phases of the Chen's scheme \cite{chen}}
\label{phase}
\end{figure*}

In this section, we investigate Chen's scheme \cite{chen}. This scheme has $4$ phases: initialization, issue, subscription and hand-off. After briefly explaining the procedure of this protocol, we mention its weaknesses.\\
The Figure \ref{shekl e chen} shows the structure of general mobile pay-TV system. The Figure \ref{phase} and the Table \ref{mshes} depict the the phases of Chen's scheme and correspondence between $MS$ and $HES$es.

\begin{table}
\caption{Correspondence between $MS$ and $HES$es}
\centering
\begin{tabular}{ccc}
\hline
$U_{i}$ &  & $S$ \\
\hline
 Initialization phase & $\leftrightarrows$ & Secure channel \\
 Issue phase & $\leftrightarrows$ & Public channel \\
 Subscription phase & $\leftrightarrows$ & Public channel \\
 Hand-off phase& $\leftrightarrows$ & Public channel \\   
\hline
\end{tabular}
\label{mshes}
\end{table}


\begin{itemize}

\item \textbf{Initialization phase}\\ Users are registered in $DBS$ of $HES$ through $SAS/SMS$, and their $ID$ is stored in $DBS$ along with $N.$ If $N=0$, the user's identity and $N=1$ is stored. These communications are carried out through a secure channel. To perform this process, the following steps are performed by $U_{i}$:\\
$U_{i}$ chooses $ID_{i}$, $PW_{i}$, and generates $b$. Then it computes $PWB_{i}=h(PW_{i}\oplus b)$ and submits $ID_{i}$ and $PWB_{i}$ to $S$.\\
$S$ checks $ID_{i}$ is already in its database or not. Then it calculates: $$K=h(ID_{i}\oplus PWB_{i})$$ $$Q=h(UD\Vert x)\oplus PWB_{i}$$ $$R=h(PWB_{i}\Vert ID_{i})\oplus h(y)$$ Here $UD=h(ID_{i}\Vert N)$.
$S$ issues the smart card containing [$K,R,Q$] and sends it to $U_{i}$ over secure channel.\\
$U_{i}$ stores $b$ on smart card. The smart card contains [$K,R,Q,b$]

\item \textbf{Issue phase}\\ For each $log-in$ and acquisition of service from each $HES$ in the network, the user should send a $log-in$ request and receives a $\Theta$ as a token. This token is used in the subscription phase. Kim \textit{et al.} has attacked this phase \cite{kim}. The attack scenario is as follows:\\
$\mathcal{A}$ listens to a user's session in the issue phase and steals the user's smart card. $\mathcal{A}$ could obtain $P$ from $C_{i}=h(P\Vert CID_{i}\Vert T_{1}\Vert n_{i})$ by using the values of $C_{i}$, $CID_{i}$, $T_{1}$, and $n_{i}$ from the intercepted messages \cite{kim}. Due to the security of the secure one-way hash function, the probability of retrieving $P$ from $C_{i}$ is negligible ($\varepsilon$). As a result, $\mathcal{A}$ cannot impersonate the user, and Chen's scheme is secure against stolen smart card attacks. In the following we describe the issue phase of Chen's scheme:\\
$U_{i}$ enters $ID_{i}$ and $PW_{i}$ and computes: $$PWB_{i}=h(PW_{i}\oplus b)$$ $$K=h(ID_{i}\oplus PWB_{i})$$ $$P=Q\oplus PWB_{i}$$ $$h(y)=h(PWB_{i}\Vert ID_{i})\oplus R$$
Then it generates a random number $n_{i}$ and calculates: $$R_{t}=R_{i}\oplus h(y\Vert n_{i})$$ $$CID_{i}=ID_{i}\oplus h(y\Vert T_{1}\Vert n_{i})$$ $$C_{i}=h(P\Vert CID_{i}\Vert T_{1}\vert n_{i})$$
and sends $m=[R_{i},C_{i},CID_{i},T_{1},n_{i}]$ to $HES$.\\
$HES$ receives $m$ at $T_{2}$ and performs the following steps:\\
Checks $T_{2}-T_{1}\leq \Delta T$ (acc/rej). Then it calculates: $$ID_{i}=CID_{i}\oplus h(y\Vert T_{1}\Vert n_{i})$$ and verifies $ID_{i}$ if is a valid user identity. Else, it terminates the $log-in$ request. Then calculates: $$P'=h(UD\Vert x)=h(h(ID_{i}\Vert N)\Vert x)$$ $$C'_{i}=h(P'\Vert CID_{i}\Vert T_{1}\Vert n_{i})$$ then it checks $C'_{i}=C_{i}$, if they are equal $HES$ accepts the $log-in$ request and calculates $R_{t}=R_{i}\oplus h(y\Vert n_{i})$. Now, it chooses $\Theta _{i}$, then calculates: $$D_{i}=h(P'\Vert CID_{i}\Vert T_{2}\Vert n_{i})$$ $$E_{i}=\Theta _{i}\oplus h(P'\Vert T_{2}\Vert n_{i})$$ $HES$ broadcasts the mutual authentication message $m_{2}=[D_{i},E_{i},T_{2}]$.\\
$U_{i}$ receives $m_{2}$ and checks the $T_{3}-T_{2}\leq \Delta T$ (acc/rej). Then it computes $D'_{i}=h(P\Vert CID_{i}\Vert T_{2}\Vert n_{i})$ and checks $D'_{i}=D_{i}$. Finally, it calculates certified token $\theta =E_{i}\oplus h(P\Vert T_{2}\Vert n_{i})$ as the session key to get Pay-TV service. 

\item \textbf{Subscription phase}\\ For communicating with $HES$ using the obtained $\Theta$ communicated with $HES$ and calculated $\gamma.$ Then, it communicated with $HES$ and set the authentication key. To calculate $\gamma$ the following steps should be done:\\
$U_{i}$ inputs its $ID$ and $PW$ and computes: $$PW=h(PW_{i}\oplus b)$$ $$h(ID_{i}\oplus PWB_{i})$$ $$K=h(ID_{i}\oplus PWB_{i})$$ $$P=Q\oplus PWB_{i}$$ $$h(y)=h(PWB_{i}\Vert ID_{i})$$ then it generates a random number $n_{i}$ and calculates: $$R_{i}=\theta _{i}\oplus h(y\Vert n_{i})$$ $$CID_{i}=ID_{i}\oplus h(y\Vert T_{1}\Vert n_{i})$$ $$C_{i}=h(P\Vert CID_{i}\Vert T_{1}\Vert n_{i})$$ and sends $m=[R_{i},C_{i},CID_{i},T_{1},n_{i}]$ to $HES$.\\
$HES$ receives $m$ at $T_{2}$  and checks $T_{2}-T_{1}\leq \Delta T$ (acc/rej). Then calculates $ID_{i}=CID_{i}\oplus h(y\Vert T_{1}\Vert n_{i})$ and verifies if $ID_{i}$ is valid user's identity and computes $P'=h(UD\Vert x)=h(h(ID_{i}\Vert N)\Vert x)$. Then it checks $C'_{i}=h(P'\Vert CID_{i}\Vert T_{1}\Vert n_{i})=C_{i}$. If they are equal, $HES$ accepts the $log-in$ request and computes $\theta =R_{i}\oplus h(y\Vert n_{i})$. $HES$ chooses a random number $\gamma _{i}$ for $U_{i}$ and calculates: $$D_{i}=h(P'\Vert CID_{i} \Vert T_{2}\Vert n_{i})$$ $$E_{i}=\gamma _{i} \oplus h(P'\Vert T_{2}\Vert n_{i})$$ Finally, it broadcasts $m_{2}=[D_{i},E_{i},T_{2}]$.\\
After receiving $M_{2}$ in $T_{3}$, $U_{i}$ checks $T_{3}-T_{2}\leq \Delta T$ and if is valid checks $D'_{i}=h(P\Vert CID_{i}\Vert T_{2}\Vert n_{i})=D_{i}$. Then calculates the certified token $\gamma _{i}=E_{i}\oplus h(P\Vert T_{2}\Vert n_{i})$ to get Pay-TV service.

\item \textbf{Hand-off phase}\\ In the hand-off phase for leaving the covered area of each $HES$ and communicating with another $HES$, another $\gamma$ should be calculated as $\gamma_{i}$ and used for obtaining future services from $HES.$ In fact, in this phase, the users are re-authenticated without re$log-in$ and set a new authentication session key to obtain new $HES$. To calculate new authentication session key, $U_{i}$ should be done the following steps:\\
It generates a new random number $n_{i}$ and computes: $$Z_{i}=\theta _{i} \oplus h(y\Vert n_{i})$$ $$CID_{i}=ID_{i}\oplus h(y\Vert T_{1}\Vert n_{i})$$ $$C_{i}=h(P\Vert CID_{i}\Vert n_{i})$$ then it sends $m=[Z_{i},C_{i},CID_{i},T_{1},n_{i}]$ to $HES$\\
$HES$ receives $m$ at $T_{2}$ and checks $T_{2}-T_{1}\leq \Delta T$, then calculates: $$ID_{i}=CID_{i}\oplus h(y\Vert T_{1}\Vert n_{i})$$ $$P'=h(h(ID_{i}\Vert N)\Vert x)$$ and checks $C'_{i}=h(P'\Vert CID_{i}\Vert T_{1}\Vert n_{i})=C_{i}$ and if they are equal accepts the request. For verifying $U_{i}$'s request, it calculates $\theta _{i}=Z_{i}\oplus h(y\Vert n_{i})$ and chooses $\gamma$ as  authentication session key and calculates: $$D_{i}=h(P'\Vert CID_{i}\Vert T_{2} n_{i})$$ $$F_{i}=\gamma _{i} \oplus h(P'\Vert T_{2}\Vert n_{i})$$ and broadcasts the mutual authentication message $m_{2}=[D_{i},F_{i},T_{2}]$.\\
$U_{i}$ receives $m_{2}$ at $T_{3}$ and checks $T_{3}-T_{2}\leq \Delta T$ and $D'_{i}=h(P'\Vert CID_{i}\Vert T_{2}\Vert n_{i})=D_{i}$ and if they are equal, it accepts $HES$'s request of mutual authentication. Finally $U_{i}$ calculates $\gamma _{i}=F_{i}\oplus h(P\Vert T_{2}\Vert n_{i})$ as the authentication session key to obtain new $HES$'s service.

\end{itemize}

\subsection{The Weaknesses of Chen et al. Scheme}
In this section, we mention the weaknesses of Chen scheme, including privilege insider attack (subsequent breaking user privacy by the $HES$, means that the malicious $HES$ can obtain the users' identity and traces them), and user traceability.

\subsubsection{Privilege insider attack}
According to the section $3.2$ and $3.4$, we assume that the $HES$es are malicious. In the issue phase of Chen protocol, all $HES$es have $y$, which is the particular key of the server, and therefore they can use it to calculate $R_{t}=R_{i}\oplus h(y\Vert n_{i})$ and $ID_{i}=CID_{i}\oplus h(y\Vert T_{1}\Vert n_{i})$. In fact, is clear, each $HES$es can calculate values such as $R_{i}$ and $ID_{i}$ through $y$. On the table \ref{alg1} ($algorithm1$), we describe this process in detail.

\begin{table}
\caption{The Privilege Insider Attack}
\centering
\begin{tabular}{l}
\hline
\textbf{Algorithm1}\\
\hline
\textbf{Set up:}\\
Input: $CID_{i}$, $T_{1}$, \& $n_{i}$ received from public channel\\
Output: $ID_{i}$\\
\textbf{Challenge:}\\
1. Receives $CID_{i}$, $T_{1}$ \& $n_{i}$ from public channel\\
2. Computes $ID^{*}_{i}=CID_{i}\oplus h(y\Vert T_{1}\Vert n_{i})$\\
 \textbf{Guess:}\\
3. \textbf{If} $ID^{*}_{i}=ID_{i}$ then\\
Return $1$ and accept $ID^{*}_{i}$ as valid $ID_{i}$ (user's $ID$)\\
\textbf{else}\\
return $0$\\

\hline
\end{tabular}
\label{alg1}
\end{table}


After obtain the real users' identity, Some users' privacy are broken as fellows:

\begin{itemize}
\item \textbf{Breaking User Anonymity:}
The $ID_{i}$ is not directly located on the channel, users' $ID$s can be accessed by a simple relation using the information received from the public channel. To obtain the user's identity, it is enough to calculate the $ID_{i}= CID_{i}\oplus h(y||T_{1}||n_{i})$ via the $CID_{i}$, $n_{i}$, and $T_{i}$ received from the public channel and knowing $y$. In such a case, the user's $ID$ can be retrieved. This procedure is described in Table \ref{alg1} ($Algorithm 1$).\\
According to the Table \ref{alg1}, We prove that malicious $HES$ succeeds in the designed game. Therefore, $\mathcal{A}$ can retrieve a real user $ID$ simply with a probability of $1$, so: $Succ_{\mathcal{A}}^{Chen}=1.$
\item \textbf{User Traceability:}
According to the procedure demonstrated in Table \ref{alg1}, malicious $HES$ is able to find the user's identity easily and grabs $ID_{i}.$ Although, it can trace similarly the user with the $algorithm1$, cannot obtain that which user re-authenticates without re-login. But it can the user in the hand off phase is same user that has been in issue phase.
\end{itemize}

\subsubsection{User traceability}
According to definition of \textit{user traceability} mentioned in the section $3.2$, $\mathcal{A}$ should not get any information about users identity. It is clearly that $\mathcal{A}$ has no information about $y$. with having $y$, $\mathcal{A}$ can obtain $ID_{i}$. But, there is no need to obtain the real user identity. In this attack, $\mathcal{A}$ wants to know the authenticated user is the same user that was previously $loged-in$ to the server or not.\\ To achieves its goal, after eavesdropping each $CID_{i}$, $T_{1}$, and $n_{i}$ form public channel, $\mathcal{A}$ chooses randomly unique $y^{*}$ and calculates $ID_{i}^{*}=CID_{i}\oplus h(y^*\Vert T_{1}\Vert n_{i})$ as pseudonym of  all users send $log-in$ request and stores all calculated $ID_{i}^{*}$s in its memory. After a while, $\mathcal{A}$ eavesdrops the public channel and calculate $ID_{i}^{*}$ and compares it with stored set of $ID_{i}^{*}$s. If new calculated $ID_{i}^{*}$ is in stored set, the user is in the authentication phase, is the same user which $\mathcal{A}$ calculated its pseudonym and stored its $ID^{*}$. Now, $\mathcal{A}$ can guess this user was $loged-in$ on the server or not with probability of $1$. In fact it can classify all users in two groups. The first group: the group which has anonymous members, but $\mathcal{A}$ knows the group members was $loged-in$ on the server. The second group: it has the anonymous members, but $\mathcal{A}$ knows the groups members never $loged-in$ on the server. We describe this process on the Table \ref{alg2}.

\begin{table}
\caption{The User traceability}
\centering
\begin{tabular}{l}
\hline
\textbf{Algorithm2}\\
\hline
\textbf{Set up:}\\
Input: $CID_{i}$, $T_{1}$, \& $n_{i}$ eavesdropped from public channel\\
Output: $0$ or $1$\\
\textbf{Challenge:}\\
1. Eavesdrops $CID_{i}$, $T_{1}$, \& $n_{i}$ from public channel\\
2. Chooses randomly $y^{*}$\\
3. Computes $ID^{*}_{i}=CID_{i}\oplus h(y^{*}\Vert T_{1}\Vert n_{i})$\\
4. Creates a set of $ID_{i}^{*}$s and stores it\\
\textbf{Guess:}\\
5. Challenger has to calculate new $ID^{*}_{i}$ like line $3$ \\
\textbf{If} the new calculated $ID^{*}_{i}$ is in the created set \\
 return $1$ (success), $ID^{*}_{i}$ was $loged-in$ on the server\\
\textbf{else}\\
return $0$ (failure), $ID^{*}_{i}$ never $loged-in$ on the server\\
\hline
\end{tabular}
\label{alg2}
\end{table}


\subsubsection{Soundness}
In the issue phase, subscription phase and hand-off phase of Chen's scheme the following parameters are calculated by users: $$R_{i}=R_{t}\oplus h(y\Vert n_{i})$$ $$CID_{i}=ID_{i}\oplus h(y\Vert T_{1}\Vert n_{i})$$ $$R_{i}=\Theta _{i}\oplus h(y\Vert n_{i})$$ $$Z_{i}=\Theta _{t}\oplus h(y\Vert n_{i})$$ According to Chen's scheme, $y$ is the secret key of the remote server stored in the hash function \cite{chen}. So, the users have no ability to calculate the mentioned parameters and they cannot use $y$.

\subsection{Our Improved Scheme}
In this section, we propose an improved issue of Chen scheme. Our improved scheme has $4$ phases which we  describe as below, and compare our changes with the original scheme. We represent the protocol's procedure indetails in Tables \ref{init} to \ref{hoff}. 

\begin{itemize}

\item \textbf{Initialization Phase}\\
Which is shown in Table \ref{init}, the server calculates  $R_{i}$, and $Q_{i}$ according to the Table \ref{init} after receiving $ID_{i}$, and $PWB_{i}$, and then stores $R_{i}$, $Q_{i}$ and $PWB_{i}$ in its database. In the following, we describe this process in details:\\
$\mathbf{U_{i}}$: $U_{i}$ generates $b$ as random number and chooses $PW_{i}$. Then it computes $PWB=h(PW_{i}\Vert b)$ and sends $PWB_{i}$ and $ID_{i}$ to the pay-TV server.\\
$\mathbf{S}$: After receiving $PWB_{i}$ and $ID_{i}$, $S$ computes $Q_{i}=h(ID_{i}\Vert x)\oplus PWB_{i}$ and $R_{i}=h(PWB_{i}\Vert ID_{i})$. Then it stores $R_{i}$, $Q_{i}$, and $Q_{i}\oplus PWB_{i}$ in its database and issues smart card containing [$R_{i}$, $Q_{i}$]. $S$ send issued smart card to $U_{i}$.\\
$\mathbf{U_{i}}$: $U_{i}$ stores $b$ on smart card memory and keeps it secure.

\begin{table*}
\caption{Initialization phase - our improved scheme}
\centering
\begin{tabular}{rcl}
\hline
$U_{i}$ &  & $S$ \\
\hline
Chooses $b$ as random number and inputs $ID_{i}$, $PW_{i}$ \& $b$ &  &  \\
 Computes $PWB_{i}=h(PW_{i}\oplus b)$ & $\xrightarrow{PWB_{i}, ID_{i}}$ & Computes \\
  &  & $Q_{i}=h(ID_{i}\Vert x)\oplus PWB_{i}$ \\
  &  & $R_{i}=h(PWB_{i}\Vert ID_{i})$ \\
 Stores random number $b$ on smart card and smart &  & Stores ($Q_{i}$,$R_{i}$ \& $Q_{i}\oplus PWB_{i}$) in $DBS$ \\
 card contains [$R_{i}$, $Q_{i}$ \& $b$] & $\xleftarrow{[R_{i}, Q_{i}]}$ & Issues a smart card containing [$R_{i}$, $Q_{i}$]\\
\hline
\end{tabular}
\label{init}
\end{table*}

\item \textbf{Issue Phase}\\
The mobile user generates $n_{i}$ as random number and calculates $R_{i}$ via the $PW_{i}$, $ID_{i}$, and $R_{i}$ and being authenticated. Then, it calculates and sends a $log-in$ request to $HES$. 
In the next step, after  the time stamp and the user's $ID$ verification, the server calculates $E_{i}$ and $D_{i}$  and broadcasts $m_{2}$.
The user also calculates $\Theta$ as the Authentication session key after checking $\Delta T$ and verifying its value. This session is shown in Table \ref{iss}. The mentioned process is depicted in the following:\\
$\mathbf{U_{i}}$: $U_{i}$ inputs its $ID$, and $PW$ and computes: $$PWB_{i}=h(PW_{i}\oplus b)$$ $$R_{i}=h(PWB_{i}\Vert ID_{i})$$ Smart card checks $R_{i}$ and verifies it. Then generates $n_{i}$ and calculates: $$Kn=Q_{i}\oplus PWB_{i}\oplus n_{i}$$ $$CID_{i}=Kn\oplus h(Kn\Vert T_{1}\Vert n_{i})$$ $$C_{i}=h(Q_{i}\Vert CID_{i}\Vert T_{1}\Vert n_{i})$$ $$R_{t}=R_{i}\oplus Kn$$ Smart card sends $m_{1}=[Kn, C_{i}, T_{1}, n_{i}]$ to $S$.\\
$\mathbf{S}$: $S$ receives $m_{1}$ at $T_{2}$ and checks $T_{2}-T_{1}\leq \Delta T$. Then it computes $Kn\oplus n_{i}=Q_{i}\oplus PWB_{i}$ and searches it in its database, then verifies it. Else, terminates this phase. $S$ checks $C'_{i}=h(Q_{i}\Vert Kn \oplus h(Kn\Vert T_{1}\Vert n_{i})\Vert T_{1}\Vert n_{i})=C_{i}$ and verifies it. Then it chooses the token $\Theta$ and stores it on $DBS$  and computes: $$D_{i}=h(R_{i}\oplus Kn \Vert CID_{i}\Vert T_{2}\Vert n_{i})$$ $$E_{i}=\Theta \oplus h(Q_{i}\Vert T_{2}\Vert n_{i}\oplus Q_{i}\oplus Kn)$$ and broadcasts $m_{2}=[D_{i}, E_{i}, T_{2}]$.\\
$\mathbf{U_{i}}$: $U_{i}$ receives $m_{2}$ at $T_{3}$  and checks $T_{3}-T_{2}\leq \Delta T$. Then it checks $D'_{i}=h(R_{t}\Vert CID_{i}\Vert T_{2}\Vert n_{i})=D_{i}$ and calculates $\Theta=E_{i}\oplus h(Q_{i}\Vert T_{2}\Vert n_{i}\oplus PWB_{i})$ as authentication session key.


\begin{table*}
\caption{Issue phase - our improved scheme}
\centering
\resizebox{\textwidth}{!}{
\begin{tabular}{rcl}
\hline
$U_{i}$ [$R_{i}$, $Q_{i}$ \& $b$] &  & $S$ [$Q_{i}, R_{i}, Q_{i}\oplus PWB_{i}$] \\
\hline
 Inputs $ID_{i}$ \& $PW_{i}$ &  &  \\
 Computes $PWB_{i}=h(PW_{i}\oplus b)$ &  &  \\
 Verifies $R_{i}=h(PWB_{i}\Vert ID_{i})$ (acc/rej) &  &  \\
 Generates $n_{i}$ and computes $Kn=Q_{i}\oplus PWB_{i}\oplus n_{i}$ &  &  \\
 Computes $CID_{i}=Kn\oplus h(Kn\Vert T_{1}\Vert n_{i})$ &  & Receives message at $T_{2}$ \\
 $C_{i}=h(Q_{i}\Vert CID_{i}\Vert T_{1}\Vert n_{i})$ & $\xrightarrow{m_{1}=[Kn, C_{i}, T_{1}, n_{i}]}$ & Checks $T_{2}-T_{1}\leq \Delta T$ \\
 $R_{t}=R_{i}\oplus Kn$ &  & Computes $Kn\oplus n_{i}=Q_{i}\oplus PWB_{i}$ \\
  &  & Checks $Q_{i}\oplus PWB_{i}$ (acc/rej) \\
  &  & $C'_{i}=h(Q_{i}\Vert Kn\oplus h(Kn\Vert T_{1}\Vert n_{i}) \Vert T_{1}\Vert n_{i})$ (acc/rej) \\
 Receives $m_{2}$ at $T_{3}$ \& checks $T_{3}-T_{2}\leq\Delta T$ & $\xleftarrow{m_{2}=[D_{i},E_{i}, T_{2}]}$ & Chooses the token $\Theta$ \& store in DBS \\
 Computes $D'_{i}=h(R_{t}\Vert CID_{i}\Vert T_{2}\Vert n_{i})$ (acc/rej)&  & Computes $D_{i}=h(R_{i}\oplus Kn\Vert CID_{i}\Vert T_{2}\Vert n_{i})$ \\
 \textbf{Authentication session Key} $\Theta =E_{i}\oplus h(Q_{i}\Vert T_{2}\Vert n_{i}\oplus PWB_{i})$ &  & $E_{i}=\Theta \oplus h(Q_{i}\Vert T_{2}\Vert n_{i}\oplus Q_{i}\oplus Kn)$  \\
\hline
\end{tabular}
}
\label{iss}
\end{table*}


\item \textbf{Subscription Phase}\\
Which is shown in Table \ref{subs}. After $\Theta$ calculation and entering $PW_{i}$ and $ID_{i}$, the user sends $Kn^{new}$, $C_{i}$ and $CID_{i}$ using the obtained $\Theta$ along with $n_{i}^{new}$ and $T_{1}$ to $HES.$ If $HES$ authenticates the user's $ID$, it will broadcast $m_{2}=[D_{i}, E_{i}, T_{2}]$ which contains $\gamma$. In the following, We describe the subscription phase of our alleviated scheme in details:\\
$\mathbf{U_{i}}$: $U_{i}$ inputs its $ID$, and $PW$ and computes $PWB_{i}=h(PW_{i}\oplus b)$. Then verifies $R_{i}=h(PWB_{i}\Vert ID_{i})$ and generates $n_{i}^{new}$ and calculates following parameters: $$Kn^{new}=Q_{i}\oplus PWB_{i}\oplus n_{i}^{new}$$ $$CID_{i}=Kn^{new}\oplus h(Kn^{new}\Vert T_{1}\Vert n_{i}^{new})$$ $$C_{i}=h(Q_{i}\Vert CID_{i}\Vert T_{1}\Vert n_{i}^{new})$$ $$R_{i}=\Theta \oplus Kn^{new}$$ $U_{i}$ sends $m_{1}=[Kn^{new}, C_{i}, T_{1}, n_{i}^{new}]$ to $HES$.\\
$\mathbf{S}$: $S$ receives $m_{1}$ at $T_{2}$ and checks $T_{2}-T_{1}\leq \Delta T$. Then computes $Kn^{new}\oplus n_{i}^{new}=Q_{i}\oplus PWB_{i}$ and checks $C'_{i}=h(Q_{i}\Vert CID_{i}\Vert T_{1}\Vert n_{i}^{new})=C_{i}$. $S$ calculates $\Theta =R_{i}\oplus Kn^{new}$ and chooses $\gamma$ as token for $U_{i}$. $S$ computes: $$D_{i}^{new}=h(R_{i}\Vert CID_{i}\Vert T_{2}\Vert n_{i}^{new})$$ $$E_{i}^{new}=\gamma _{i}\oplus h(R_{i}\Vert T_{2}\Vert n_{i}^{new}\oplus Q_{i}\oplus Kn^{new})$$ $S$ broadcasts $m_{2}=[D_{i}^{new}, E_{i}^{new}, T_{2}]$.\\
$\mathbf{U_{i}}$: receives $m_{2}$ at $T_{3}$ and checks $T_{3}-T_{2}\leq \Delta T$. Then it checks  $D_{i}^{'new}=h(R_{i}\Vert CID_{i}\Vert T_{2}\Vert n_{i}^{new})=D_{i}^{new}$ and computes $\gamma _{i}=E_{i}^{new}\oplus h(R_{i}\Vert T_{2}\Vert n_{i}^{new}\oplus PWB_{i})$ as authentication session key to get services.


\begin{table*}[t]
\caption{Subscription phase - our improved scheme}
\centering
\resizebox{\textwidth}{!}{
\begin{tabular}{rcl}
\hline
$U_{i}$ [$R_{i}$, $Q_{i}$ \& $b$] &  & $S$ [$Q_{i}, R_{i}, Q_{i}\oplus PWB_{i}$] \\
\hline
 Inputs $ID_{i}$ \& $PW_{i}$ &  &  \\
 Computes $PWB_{i}=h(PW_{i}\oplus b)$ &  &  \\
 Verifies $R_{i}=h(PWB_{i}\Vert ID_{i})$ (acc/rej) &  &  \\
 Generates $n_{i}^{new}$ and computes $Kn^{new}=Q_{i}\oplus PWB_{i}\oplus n_{i}^{new}$ &  &  \\
 Computes $CID_{i}=Kn^{new}\oplus h(Kn^{new}\Vert T_{1}\Vert n_{i}^{new})$ &  & Receives message at $T_{2}$ \\
 $C_{i}=h(Q_{i}\Vert CID_{i}\Vert T_{1}\Vert n_{i}^{new})$ & $\xrightarrow{m_{1}=[Kn^{new}, C_{i}, T_{1}, n_{i}^{new}]}$ & Checks $T_{2}-T_{1}\leq \Delta T$ \\
 $R_{i}=\Theta\oplus Kn^{new}$ &  & Computes $Kn^{new}\oplus n_{i}^{new}=Q_{i}\oplus PWB_{i}$ \\
  &  & Checks $Q_{i}\oplus PWB_{i}\oplus h(y\Vert R_{i})$ (acc/rej) \\
  &  & $C'_{i}=h(Q_{i}\Vert Kn^{new}\oplus h(Kn^{new}\Vert T_{1}\Vert n_{i}^{new})\Vert T_{1}\Vert n_{i}^{new})$  (acc/rej)\\
 Receives $m_{2}$ at $T_{3}$ \& checks $T_{3}-T_{2}\leq\Delta T$ & $\xleftarrow{m_{2}=[D_{i}^{new},E_{i}^{new}, T_{2}]}$ & $\Theta =R_{i}\oplus Kn^{new}$ \\
 Computes $D'^{new}_{i}=h(R_{i}\Vert CID_{i}\Vert T_{2}\Vert n_{i}^{new})$ (acc/rej) &  & Chooses token $\gamma$ for $U_{i}$ \\
 Computes $\gamma _{i}=E_{i}^{new}\oplus h(R_{i}\Vert T_{2}\Vert n_{i}^{new}\oplus PWB_{i})$ as  &  & Computes $D_{i}^{new}=h(R_{i}\Vert CID_{i}\Vert T_{2}\Vert n_{i}^{new})$  \\
 \textbf{Authentication key} to get services &  & $E_{i}^{new}=\gamma _{i}\oplus h(R_{i}\Vert T_{2}\Vert n_{i}^{new}\oplus Q_{i}\oplus Kn^{new})$  \\
\hline
\end{tabular}
}
\label{subs}
\end{table*}


\item \textbf{Hand-off Phase}\\
Any user who wants to leave a $HES$ region and $log-in$ to another $HES$ region have to go through this step according to Table \ref{hoff}. Since the user is in the primary $HES$ region, no $log-in$ is required. In fact, the user is re-authenticated without re-login. When this step is finished, the user obtains $\gamma _{i}^{new}$ for communicating with the new $HES$. The hand-off phase of our alleviated scheme is shown on the Table \ref{hoff}.\\
According to the Table \ref{hoff}, $S$ replaces $PWB_{i}\oplus Q_{i}\oplus h(y\Vert R_{i})$ on $PWB_{i}\oplus Q_{i}$ stored in its database.

\begin{table*}
\caption{Hand-off phase - our improved scheme}
\centering
\resizebox{\textwidth}{!}{
\begin{tabular}{rcl}
\hline
\multicolumn{3}{c}{Re-authentication without re-login}\\
$U_{i}$ [$R_{i}$, $Q_{i}$ \& $b$] &  & $S$ [$Q_{i}, R_{i}, Q_{i}\oplus PWB_{i}$] \\
\hline
 Generates $n_{i}^{new}$ and computes $Kn^{new}=Q_{i}\oplus PWB_{i}\oplus n_{i}^{new}$ &  &  \\
 Computes $CID_{i}=Kn^{new}\oplus h(Kn^{new}\Vert T_{1}\Vert n_{i}^{new})$ &  & Receives message at $T_{2}$ \\
 $C_{i}=h(Q_{i}\Vert CID_{i}\Vert T_{1}\Vert n_{i}^{new})$ & $\xrightarrow{m_{1}=[C_{i}, T_{i},n_{i}^{new}]}$ & Checks $T_{2}-T_{1}\leq \Delta T$ \\
  &  & Computes $Kn^{new}\oplus n_{i}^{new}=Q_{i}\oplus PWB_{i}$ \\
  &  & Checks $Q_{i}\oplus PWB_{i}\oplus h(y\Vert R_{i})$ (acc/rej) \\
  & & Replaces $PWB_{i}\oplus Q_{i}\oplus h(y\Vert R_{i})$ on $PWB_{i}\oplus Q_{i}$ \\
  &  & $C'_{i}=h(Q_{i}\Vert Kn^{new}\oplus h(Kn^{new}\Vert T_{1}\Vert n_{i}^{new})\Vert T_{1}\Vert n_{i}^{new})$ (acc/rej) \\
  &  & Chooses the new authentication session key $\gamma ^{new}$ and \\
Receives $m_{2}$ at $T_{3}$ and checks $T_{3}-T_{2}\leq \Delta T$ (acc/rej) & $\xleftarrow{m_{2}=[D_{i}^{new},F_{i}, T_{2}]}$ & computes $D_{i}^{new}=h(R_{i}\Vert CID_{i}\Vert T_{2}\Vert n_{i}^{new})$ \\
Computes $D^{'new}_{i}=h(R_{i}\Vert CID_{i}\Vert T_{2}\Vert n_{i}^{new})$ &  & $F_{i}=\gamma _{i}^{new} \oplus h(R_{i}\Vert T_{2}\Vert n_{i}^{new}\oplus Q_{i}\oplus Kn^{new})$ \\
Computes $\gamma _{i}^{new}=F_{i}\oplus h(R_{i}\Vert T_{2} \Vert n_{i}^{new}\oplus PWB_{i})$ &  &  \\
as \textbf{new Authentication session key} to obtain new HES &  &  \\
\hline
\end{tabular}
}
\label{hoff}
\end{table*}

\end{itemize}

\subsection{Security Analysis of Our Improved Scheme}
This section is composed of three subsections. After explanation of the reason of our changes, analyze the improved scheme both heuristically and formally is analyzed.\\
Now, we analyze the main changes in our improved scheme compared with Chen's scheme. depicted in Tables \ref{init} to \ref{hoff}.

\begin{itemize}

\item \textbf{Removing $N$ from DBS of $HES$:}  By storing $Q_{i}$, $R_{i}$, and  $PWB_{i}\oplus Q_{i}$, the server does not need to store $N$ anymore in $DBS$ of $HES$. Each user authenticates anonymously after sending the $log-in$ request for each $HES$ in the authentication phase by $HES$ with stored parameters in $DBS$ of $HES$.
\item \textbf{Removing $h(y)$  from $R$:} As in the Chen scheme, $h(y)$  is the public key of the server, and it is available to all users. Its presence or absence in $R$ value does not guarantee any security.
\item \textbf{Lack of using $y$ in the generation phase:} We do not use $y$ to prevent "user impersonation" and "user traceability" attacks. We used $Q_{i}$, and $R_{i}$ instead. $Q_{i}$ and $R_{i}$ are joint parameter between $U_{i}$ and $HES$. $R_{i}$ and $Q_{i}$ are stored in the user memory and $DBS$ of  $HES$ produced by the server.
\end{itemize}

\subsubsection{Heuristic Security Analysis}

In this section, we analyze the improved scheme in heuristic form and show that our scheme is resistant to all prevalent attacks. Imagin $\mathcal{A}$ has possession of sensetive information stored on the card (with power attack \cite{chit}). we prove that the scheme resists a stolen smart card or stolen server. So, $\mathcal{A}$ cannot evade the users' privacy or create any interference in communications. 

\begin{itemize}

\item \textbf{Stolen server database attack and stolen verification table attack:}
By stealing the information stored on the server, $\mathcal{A}$ achieves $R_{i}=h(PWB_{i}\Vert ID_{i})$, $Q_{i}=h(ID_{i}\Vert x)\oplus PWB_{i}$ ,$PWB_{i}=h(PW_{i}\oplus b)$. We proof that $\mathcal{A}$ has no ability to obtain sensitive parameters:
\begin{itemize}
\item For $x$ retrieval, $\mathcal{A}$ needs to retrieve the hash function value, which is impossible given the secure $hash$ function. Therefore, $\mathcal{A}$ must again try to retrieve $ID_{i}$ first. Then it can run brute force attack on $x$. So, its success probability is $(1/2)^{(Length-of-x)+(Length-of-ID_{i})}$.
\item For $PW_{i}$ retrieval, $\mathcal{A}$ needs to retrieve the hash function value, which is impossible given the secure $hash$ function. Therefore, $\mathcal{A}$ must again try to retrieve $b$ first. Then, it can retrieval $PW_{i}$. So, its success probability is $(1/2)^{(Length-of-b)+(Length-of-PW_{i})}$.
\item The other parameter which $\mathcal{A}$ wants to retrieve it, is $ID_{i}$ stored in hash function. For retrieval it, $\mathcal{A}$ has to run brute force attack with probability of $(1/2)^{Length-of-ID_{i}}$.
\end{itemize}

\item \textbf{Stolen smart card attack:}
By server stealing and after power analysis, $\mathcal{A}$ achieves $Q_{i}=h(ID_{i}\Vert x)\oplus PWB_{i}$, $R_{i}=h(PWB_{i}\Vert ID_{i})$, and random number $b$. $b$ does not help $\mathcal{A}$ to obtain the sensitive information of terms $R_{i}$ and $Q$, $\mathcal{A}$ also needs an $exponential-time$, to achieve them. 

\item \textbf{Replay attack:}
There is the freshness of all sent flows on the channel and the new random number and time stamp, so there is no possibility for this attack. In fact, if $\mathcal{A}$ intends to resend the previous messages, it needs to access the term $R_{i}$. As mentioned in the previous section, exponential time is needed to produce these parameters. It should be noted that $\mathcal{A}$ could access $R_{i}$ by possessing the server, but if $\mathcal{A}$ is present at the highest level of attack (stolen server) and it possesses the database of server, there is no reason for the replay attack. 

\item \textbf{Impersonation attack:}
For user impersonation, $\mathcal{A}$ needs a pair of $(ID_{i}, PW_{i})$ or it should be able to produce $Kn$, $C_{i}$, and $CID_{i}.$ As explained in the previous sections, in order to acquire or produce the desired parameters, $\mathcal{A}$ needs exponential time and it cannot implement the attack in polynomial time. 

\item \textbf{Breaking user anonymity and user traceability attacks:}
According to the Table \ref{subs}, it is clear that no user $ID$ trace is placed directly on the channel. The only place that the user $ID$ has been used is $CID_{i}=Kn\oplus h(Kn\Vert T_{1}\Vert n_{i})$ $=Q_{i}\oplus PWB_{i}\oplus n_{i}\oplus h(Q_{i}\oplus PWB_{i}\oplus n_{i}\Vert T_{1}\Vert n_{i})$ $=h(ID_{i}\Vert x)\oplus h(PW_{i}\oplus b)\oplus h(PW_{i}\oplus b)\oplus n_{i}\oplus h(h(ID_{i}\Vert x)\oplus h(PW_{i}\oplus b)\oplus h(PW_{i}\oplus b)\oplus n_{i}\Vert T_{1}\Vert n_{i})$ that $\mathcal{A}$ is faced with this phrase with the possibility of $(1/2)^{ Length-of-hash}$ to retrieve the user $ID.$ If $\mathcal{A}$ possesses the database of a server, it can access the user's $ID$, but having the user $ID$ without any adverse information is not sufficiently useful. Card stealing and the card data retrieval do not help $\mathcal{A}$ to achieve user's $ID$s. Regarding to protection of user $ID$, the user is untraceable. $\mathcal{A}$ cannot calculate the user $ID$, so it cannot trace the user in the hand-off phase.

\item \textbf{Channel eavesdropping attack:}
According to the description in previous sections, $\mathcal{A}$ cannot actively attack by channel avoidance and having the transmitted information on the channel. Also $\mathcal{A}$ cannot able to  obtain user's $ID$ via passive attack.

\end{itemize}

\subsubsection{Formal Security Analysis}
In this section, we analyze our scheme in the formal model \cite{dabt,yand}, which is shown in Tables \ref{alg3} to \ref{alg6} ($Algorithms3$ to $Algorithms6$). In the $algorithms$, we show that our alleviated scheme is resistant against "channel eavesdrop" and "stolen card attack" and in random oracle model. By regarding the one-way hash function (Note that the parameters represented by $*$ are generated by $\mathcal{A}$).

\begin{itemize}

\item \textbf{Channel eavesdropping}\\
$\mathcal{A}$ obtains $Kn$, $C_{i}$, $T_{1}$, and $n_{i}$  with interception. To recover the sensitive information about $U_{i}$, there must be a process in accordance with the $Algorithm3$ that shown on the Table \ref{alg3}.

\begin{table}[t]
\caption{Channel Eavesdropping}
\centering
\begin{tabular}{l}
\hline
\textbf{Algorithm3} $EXP_{Imp Chen}^{Hash}$\\
\hline
\textbf{Set up:}\\
Input: $Kn$, $C_{i}$, $T_{1}$, and $n_{i}$ eavesdropped from public channel\\
Output: $1$ (success) / $0$ (failure) \\
\textbf{Challenge:}\\
1.	Eavesdrops $Kn$, $C_{i}$, $T_{1}$, and $n_{i}$ from public channel eavesdropping\\
2.	Computes $Q_{i}\oplus PWB_{i}=Kn\oplus n_{i}$ \\
where $Q_{i}=h(ID_{i}\Vert x)\oplus h(PW_{i}\oplus b)$ and  $PWB_{i}=h(PW_{i}\oplus b)$\\
3.	Selects randomly $ID_{i}^{*}$, $x^{*}$, $PW_{i}^{*}$, and $b^{*}$ \\
4. Computes $Kn^{*}\oplus n_{i}^{*}=h(ID_{i}^{*}\Vert x^{*})\oplus h(PW_{i}^{*}\oplus b^{*})\oplus h(PW_{i}^{*}\oplus b^{*})$ \\
\textbf{Guess:}\\
5.	\textbf{If} $Kn^{*}\oplus n_{i}^{*}=Kn\oplus n_{i}$ \\
Accepts selected $ID_{i}^{*}$, $x^{*}$, $PW_{i}^{*}$, and $b^{*}$ as $ID_{i}$, $x$, $PW_{i}$, and $b$ \\
Return $1$ (success)\\
\textbf{else}\\
Return $0$ (failure)\\
\hline
\end{tabular}
\label{alg3}
\end{table}

In the designed game noted on the Table \ref{alg3}, $\mathcal{A}$ eavesdrops $Kn$, $C_{i}$, $T_{1}$, and $n_{i}$ from public channel and tries to guess sensitive information. To pass the game successfully, it has to guess $ID_{i}$, $x$, $PW_{i}$, and $b$ correctly. Since the maximum probability of success is $(1/2)^{inputes-length}$. So, $\mathcal{A}$ is not able to guess sensitive information correctly in polynomial time. 
$$Succ_{\mathcal{A}-ID_{i}}^{Imp-Chen}=Pr[EXP_{\mathcal{A}}^{hash}]\leq (\dfrac{1}{2})^{ID_{i}-length}$$
$$Succ_{\mathcal{A}-x}^{Imp-Chen}=Pr[EXP_{\mathcal{A}}^{hash}]\leq (\dfrac{1}{2})^{x-length}$$
$$Succ_{\mathcal{A}-PW_{i}}^{Imp-Chen}=Pr[EXP_{\mathcal{A}}^{hash}]\leq (\dfrac{1}{2})^{PW_{i}-length}$$
$$Succ_{\mathcal{A}-b}^{Imp-Chen}=Pr[EXP_{\mathcal{A}}^{hash}]\leq (\dfrac{1}{2})^{b-length}$$
$$\mathbf{Succ_{\mathcal{A}}^{Imp-Chen}=Pr[EXP_{\mathcal{A}}^{hash}]\leq (\dfrac{1}{2})^{(ID_{i}\Vert x\Vert PW_{i}\Vert b)-length}}\leq \epsilon $$
So, it can not guess mentioned parameters correctly.

\item \textbf{Stolen smart card attack}\\
If the smart card is stolen and corrupted by power analysis attack, $\mathcal{A}$ acquires stored data and tries to impersonate the user or deceive the server. $\mathcal{A}$ this process according to Tables \ref{alg4} and \ref{alg5} ($algorithm$s $4$ and $5$). In this section, the invader tries to recover $4$ parameters. To indicate that our improved scheme is secure against this attack, we design two games shown in the Tables \ref{alg4} and \ref{alg5}. In the games, $\mathcal{A}$ obtains $R_{i}$, $Q_{i}$, and $b$ from smart card memory by power attack and runs the games mentioned in the $algorithm$ $4$ and $5$.

\begin{itemize}

\item After recovering $R_{i}$, and $b$, $\mathcal{A}$ tries to obtain the user's private key and $ID$ which are described in Table \ref{alg4} indetails.\\

\begin{table}[t]
\caption{Stolen Smart Card}
\centering
\begin{tabular}{l}
\hline
\textbf{Algorithm4} $EXP_{Imp Chen}^{Hash}$ \\
\hline
\textbf{Set up:}\\
Input: $R_{i}$, and $b$ recovered from smart card memory by power attack\\
Output: $1$ (success) / $0$ (failure) \\
\textbf{Challenge:}\\
1.	Recovers $R_{i}$, and $b$ from smart card by power analisys attack\\
2.	Selects randomly $PW_{i}^{*}$  and $ID_{i}^{*}$ as user's private key and $ID$  \\
3.	Computes $R_{i}^{*}=h(h(PW_{i}^{*}\oplus b)\Vert ID_{i}^{*})$ \\  
\textbf{Guess:}\\
4.	\textbf{If} $R_{i}^{*}=R_{i}$ then\\
Accepts $PW_{i}^{*}$ and $ID_{i}^{*}$ as user's private key and $ID$ \\
Return $1$ (success)\\
\textbf{else}\\
Return $0$ (failure)\\
\hline
\end{tabular}
\label{alg4}
\end{table}
According to the Table \ref{alg4}, $\mathcal{A}$ has no chance to obtain a user's private key and $ID$, so:
$$Succ_{\mathcal{A}-PW_{i}}^{Imp-Chen}=Pr[EXP_{\mathcal{A}}^{hash}]\leq (\dfrac{1}{2})^{PW_{i}-length}$$
$$Succ_{\mathcal{A}-ID_{i}}^{Imp-Chen}=Pr[EXP_{\mathcal{A}}^{hash}]\leq (\dfrac{1}{2})^{ID_{i}-length}$$
$$\mathbf{Succ_{\mathcal{A}}^{Imp-Chen}=Pr[EXP_{\mathcal{A}}^{hash}]\leq (\dfrac{1}{2})^{(ID_{i}\Vert PW_{i})-length}}\leq \epsilon $$

\item According to the recovering the server's private key which is described in Table \ref{alg3}, if an output of the $algorithm3$ is $1$ (but, we prove formally, its probability is negligible). It means  we assume that $\mathcal{A}$ obtains the $x$, and $PW_{i}$ successfully and tries to guess the $ID_{i}$. To proof formally that $\mathcal{A}$ has no ability to obtain the user $ID$ and breaks the user privacy, we design a game and depict it on the Table \ref{alg5}. 

\begin{table}[t]
\caption{Stolen Smart Card}
\centering
\resizebox{\linewidth}{!}{
\begin{tabular}{l}
\hline
\textbf{Algorithm5}  $EXP_{Imp Chen}^{Hash}$\\
\hline
\textbf{Set up:}\\
Input: $Q_{i}$ and $b$ recovered from smart card memory by power attack,\\ correct $x$, and $PW_{i}$ which are guess successfully from Table \ref{alg3}\\
Output: $1$ (success) / $0$ (failure) \\
\textbf{Challenge:}\\
1. Recovers $Q_{i}$ from smart card by power attack\\
2. Assumes that $x^{*}$, and $PW_{i}^{*}$ are correct \\and $Pr[x^{*}=x \cap PW_{i}^{*}=PW_{i}]=1$ \\
3. Selects randomly $ID_{i}^{*}$\\
4. Computes $Q_{i}^{*}=h(ID_{i}^{*}\Vert x)\Vert h(PW_{i}\oplus b)$\\
\textbf{Guess:}\\
\textbf{If} $Q_{i}^{*}=Q_{i}$ then\\
Return $1$ and accept $ID_{i}^{*}$ as user $ID$\\
\textbf{else}\\
return $0$\\
\hline
\end{tabular}
}
\label{alg5}
\end{table}

In the $algorithm5$ we assume that $Pr[x^{*}=x \cap PW_{i}^{*}=PW_{i}]=1$. However, $\mathcal{A}$ can obtain the user $ID$ with negligible probability. We note that: $$Succ_{\mathcal{A}}^{Imp-Chen}=Pr[EXP_{\mathcal{A}}^{hash}\vert Pr[x^{*}=x \cap PW_{i}^{*}=PW_{i}]=1]\leq \epsilon$$ We know that $Pr[x^{*}=x \cap PW_{i}^{*}=PW_{i}]\leq \epsilon$. So, $\mathcal{A}$ has no chance to guess $ID_{i}$ successfully. It is great anonymity level ($\mathcal{A}$ has the server's secret and user's password, but it cannot break the user privacy and obtain user's $ID$).

\end{itemize}

\item \textbf{Stolen server attack}\\
According our alleviated scheme, $R_{i}$, and $Q_{i}$ are parameters stored in user memory and server database. So, the designed games in this item are similar to previous item (stolen smart card attack) and there is no need to repeat the formal analysis for this item.

\item \textbf{Privilege insider attack}\\
According to the section $3.2$ and the definition of privilege insider attack, we assume that the servers are malicious and we want that they have no information about user identity. To prove that our alleviated scheme is resists to privilege insider attack, we design a game and show it on the Table \ref{alg6}.

\begin{table}[t]
\caption{Privilege insider attack}
\centering
\begin{tabular}{l}
\hline
\textbf{Algorithm6}  $EXP_{Imp Chen}^{Hash}$\\
\hline
\textbf{Set up:}\\
Input: $Kn$, $C_{i}$, $T_{1}$, and $n_{i}$ received form public channel \\
Output: $1$ (success) / $0$ (failure) \\
\textbf{Challenge:}\\
1. Receives $Kn$, $C_{i}$, $T_{1}$, and $n_{i}$ form public channel \\
2. Computes $Kn\oplus n_{i}=Q_{i}\oplus PWB_{i}$\\
3. Searches $Q_{i}\oplus PWB_{i}$ and find $Q_{i}$, and $R_{i}$\\
4. Computes $PWB_{i}=Q_{i}\oplus PWB_{i}\oplus Q_{i}$\\
(Now, the challenger has $Kn$, $C_{i}$, $T_{1}$, $n_{i}$, $Q_{i}$, $R_{i}$, and $PWB_{i}$)\\
\textbf{Note:} Challenger wants to obtain $ID_{i}$ and it uses the parameters\\ contain user $ID$\\
\textbf{Result:} Challenger uses $R_{i}$, and $Q_{i}$ to recover $ID_{i}$\\ 
5. Selects randomly $ID_{i}^{*}$\\
6. Computes $Q_{i}^{*}=h(ID_{i}^{*}\Vert x)\oplus PWB_{i}$ and\\ $R_{i}^{*}=h(PWB_{i}\Vert ID_{i}^{*})$\\
\textbf{Guess:}\\
\textbf{If} $Q_{i}^{*}=Q_{i}$, or $R_{i}^{*}=R_{i}$ \\
Return $1$ and accepts $ID_{i}^{*}$ as user $ID$\\
\textbf{else}\\
return $0$\\
\hline
\end{tabular}
\label{alg6}
\end{table}

We assumed that, we have secure one-way hash function and the success probability of obtain the hash argument is negligible. So, we note for this item:
$$Succ_{\mathcal{A}}^{Imp-Chen}=Pr[EXP_{\mathcal{A}}^{hash}]\leq (\dfrac{1}{2})^{ID_{i}-length}$$

\end{itemize}

\section{Comparison}

\begin{table*}
\caption{Security Features Comparison}
\centering
\begin{tabular}{|>{\centering}m{5cm}|>{\centering}m{0.8cm}|>{\centering}m{0.8cm}|>{\centering}m{0.8cm}|>{\centering}m{0.8cm}|>{\centering}m{0.8cm}|>{\centering}m{0.8cm}|>{\centering}m{0.8cm}|}
\hline
\backslashbox{\textbf{Scheme}}{\textbf{Security Feature}} & S1 & S2 & S3 & S4 & S5 & S6 & S7
\tabularnewline \hline Chen \textit{et al} \cite{chen} (2011) & \ding{51} & \ding{51} & \ding{51} & \ding{53} & \ding{53} &  \ding{51}  & \ding{53} 
\tabularnewline \hline Kim \textit{et al}   \cite{kim} (2012) & \ding{51} & \ding{51} & \ding{51} & \ding{53} & \ding{53} &  \ding{51}  & \ding{51}  
\tabularnewline \hline Yeh L \cite{loya} (2012) & \ding{53} & \ding{53} & \ding{51} & \ding{51} & \ding{51} &     \ding{51} & \ding{51}
\tabularnewline \hline Sabzinejad Farash \cite{mosa} (2016) & \ding{51} & \ding{51} & \ding{51} & \ding{51} & \ding{51} & \ding{53} & \ding{51} 
\tabularnewline \hline Our Improved Scheme & \ding{51} & \ding{51} & \ding{51} & \ding{51} &\ding{51} &   \ding{51} &  \ding{51}
\tabularnewline \hline \multicolumn{8}{|l|}{{\textbf{Note:}}}
\tabularnewline  \multicolumn{8}{|l|}{S1: Resistance against stolen/lost smart card attack and user  impersonation attack}
\tabularnewline  \multicolumn{8}{|l|}{S2: Resistance against stolen verifier or stolen verification table and server impersonation attack}
\tabularnewline  \multicolumn{8}{|l|}{ S3: Resistance against DoA attack}
\tabularnewline  \multicolumn{8}{|l|}{ S4: Secure against privacy preserving (compromise user's $ID$ for other server, which is not }
\tabularnewline  \multicolumn{8}{|l|}{ server that submitted user)}
\tabularnewline  \multicolumn{8}{|l|}{ S5: Secure against user traceability}
\tabularnewline  \multicolumn{8}{|l|}{ S6: Provide mutual authentication}
\tabularnewline  \multicolumn{8}{|l|}{ S7: Resistance against privilege insider attack}
\tabularnewline \hline
\end{tabular}
\label{comp}
\end{table*}

\begin{table}
\caption{Performance Comparison}
\resizebox{\linewidth}{!}{
\centering
\begin{tabular}{|c|c|c|c|c|c|c|c|c|}
\hline
\backslashbox{\textbf{Scheme}}{\textbf{Feature}} & P1 & P2 & P3 & P4 & P5 & P6 & P7 & P8 \\ 
\hline 
Chen \cite{chen} & 6 & 0.78 & 7 & 0.91 & 7 & 0.91 &  4 & 5 \\ \hline
Kim \cite{kim} & 7 & 0.91 & 20 & 1.3 & 7 & 0.91 & 4 & 5 \\ \hline
Li \cite{xili} & 5 & 0.65 & 9 & 1.17 & 20 & 2.6 & 6 & 4 \\ \hline
Our scheme & 3 & 0.39 & 6 & 0.78 & 4 & 0.52 & 3 & 4 \\ \hline
\multicolumn{9}{|l|}{\textbf{Note:}}\\
\multicolumn{9}{|l|}{P1: The number of hash function in registration phase}\\
\multicolumn{9}{|l|}{P2: The execution time of registration phase ($\mu s$)}\\
\multicolumn{9}{|l|}{P3: The number of hash function in issue phase - user side}\\
\multicolumn{9}{|l|}{P4: The execution time of issue phase - user side ($\mu s$)}\\
\multicolumn{9}{|l|}{P5: The number of hash function in issue phase - server side}\\
\multicolumn{9}{|l|}{P6: The execution time of issue phase - server side ($\mu s$)}\\
\multicolumn{9}{|l|}{P7: The number of parameters stored in the smart card}\\
\multicolumn{9}{|l|}{P8: The number of parameters send on public channel in issue phase}\\

\hline
\end{tabular}
}
\label{performance}
\end{table}

In this section, we compare our improved scheme with other schemes in both security features and performance cost.

\begin{itemize}

\item \textbf{Security features}\\
Chen \textit{et al} proposed a scheme for mobile pay-TV \cite{chen}, and then Kim \textit{et al} improved it in 2012 against the \textit{stolen card attacks} \cite{kim}. However, we mentioned in the previous sections, the improvement seems to be wrong. It has weaknesses such as \textit{breaking user privacy}, user traceability and some forms of computing like Chen's scheme. In this section, we showed in the Table \ref{comp}, the benefits of our alleviated scheme compare with that of Chen \textit{et al}.\\
According to Table \ref{comp}, our improved scheme has even more security features than Sabzinejad Farash's scheme \cite{mosa} and our alleviated scheme much more lighter the Farash's scheme. Also, our  scheme is more secure than both Chen's scheme \cite{chen} and its improvement proposed in 2012 \cite{kim}. Our scheme is secure against stolen/lost smart card, impersonation, and stolen verifier attacks, but $\mathcal{A}$ can impersonate users in Yeh L's scheme \cite{loya}.

\item \textbf{Performance cost}\\

In recent years, many anonymous athentication schemes for mobil pay-TV are proposed. Some of them only use of hash function and suitable for light device. According to \cite{ar2017}, we assume the execution time of the hash function is $0.13\mu s$ and the execution time of pairing function is $17 500.354\mu s$. It is clearly that the pairing-based schemes are much heavier and slower than the schemes use only hash function (for example, the execution time of our improved scheme in issue phase is $1.3\mu s$ and the execution time of issue phase in Wu \textit{et al} \cite{AIHC1} is about $2777.357 \mu s$).  We depict in the Table \ref{performance} performance comparison of our improved scheme with other light schemes.\\

\end{itemize}

\section{Conclusion}
To save energy, lightweight devices have become customary. Light protocols should help them to develop.  However, we have to respect to their security and privacy policies. one important aspect of privacy is an anonymity fulfilled by anonymous authentication protocols. Recently, a lot of anonymous authentication protocols have been proposed which is based on  secure hash function or bilinear pairing transform. Hash-based protocols are lightweight and quick to run. Our alleviated protocol is more secure and lighter than mentioned protocols. Since the light devices such as sensors, smart cards, and smart phones are increasing, we predict lightweight protocols and hash-based protocols will be more popular to be paid.

\end{document}